# Combining single-molecule super-resolved localization microscopy with fluorescence polarization imaging to study cellular processes

Jack W Shepherd[1,2,3], Alex L Payne-Dwyer[1,2,3], Ji-Eun Lee[1,2], Aisha Syeda[2], and Mark C Leake[1,2]*

[1]Department of Physics, University of York, York, YO10 5DD

[2]Department of Biology, University of York, York, YO10 5DD

[3]These authors contributed equally

*To whom correspondence should be addressed. E-mail mark.leake@york.ac.uk

**Abstract**

Super-resolution microscopy has catalyzed valuable insights into the sub-cellular, mechanistic details of many different biological processes across a wide range of cell types. Fluorescence polarization spectroscopy tools have also enabled important insights into cellular processes through identifying orientational changes of biological molecules typically at an ensemble level. Here, we combine these two biophysical methodologies in a single home-made instrument to enable the simultaneous detection of orthogonal fluorescence polarization signals from single fluorescent protein molecules used as common reporters on the localization of proteins in cellular processes. These enable measurement of spatial location to a super-resolved precision better than the diffraction-limited optical resolution, as well as estimation of molecular stoichiometry based on the brightness of individual fluorophores. In this innovation we have adapted a millisecond timescale microscope used for single-molecule detection to enable splitting of fluorescence polarization emissions into two separate imaging channels for s- and p- polarization signals, which are imaged onto separate halves of the same high sensitivity back-illuminated CMOS camera detector. We applied this fluorescence polarization super-resolved imaging modality to a range of test fluorescent samples relevant to the study of biological processes, including purified monomeric green fluorescent protein, single combed DNA molecules, and protein assemblies and complexes from live Escherichia coli and Saccharomyces cerevisiae cells. Our findings are qualitative but demonstrate promise in showing how fluorescence polarization and super-resolved localization microscopy can be combined on the same sample to enable simultaneous measurements of polarization and stoichiometry of tracked molecular complexes, as well as the translational diffusion coefficient.

**Introduction**

Understanding the 'physics of life' at the molecular level [1] has undergone a revolution since the development and refinement of fluorescence microscopy [2] and is now routinely used at milliseconds to a few tens of milliseconds timescales to understand the spatial organization within living cells as well as the dynamical properties of constituent biomolecules [3]. In particular, the mobility – both translational and rotational - of these biomolecules gives information on their local cellular environment and their functional interactions, i.e., *which* molecules interact with other molecules inside cells, *where* they interact in the context of their sub-cellular location, and *when* they interact in the context of past and future biological events. While translational diffusion coefficients [4] or molecular dynamics simulations [5] can be used to estimate rotational properties of biomolecules, intracellular environments are crowded, with values of ~20% (w/v) protein content or more being typical [6], and present complex diffusive landscapes in which states of rotational and translational mobility are not necessarily indicative of one another [7]. More accurate experimental measurements of rotational states are desirable, especially if coupled with the simultaneous ability to resolve spatially where in a living cell these mobile biomolecules are.

Fluorescent 'reporter' probes, i.e., fluorescent tags that report on the location of specific biological molecules, remain a powerful and selective tool in this regard. As the excited molecule collapses to its ground state and radiates, the emitted photon's electric field aligns with the transition electric dipole moment of the emitting fluorophore [8], leading to a well-defined linear polarization. With dedicated optical components in a light microscope setup, one may decompose the emission of individual fluorophores into orthogonal components, and measurement of their relative intensities confers information on the fluorophore's azimuthal angle about the optical axis of the instrument.

In the cell cytoplasm and other biologically relevant systems with relatively low orientation constraints, molecular rotation typically occurs on an approximately nanosecond timescale, meaning that measurement of rotationally-coupled properties requires specialized photon counting hardware with similarly high temporal resolution [9]. This temporal accuracy, however, typically comes at the cost of poor spatial resolution, with nanosecond scale experiments usually performed at an ensemble detection scale equivalent to several thousands of molecules [10]. Where available, simultaneous high spatial and temporal resolutions enable detailed measurements of dynamical processes, such as in studies of conformational actuations of molecular motors [11] or lipid membrane components [12], though this is typically limited by a small imaging volume that precludes reconstructing an image of a biological cell on a relevant timescale. On the other hand, imaging diffusing molecules on approximately millisecond timescales - at the sensitivity or readout limits of current fluorescence microscope cameras [13], [14] for fully sampled multi-micron fields of view - allows the molecule of interest to, potentially, rotate and tumble hundreds or thousands of times during a single frame acquisition, effectively scrambling the polarization signal [15]. For strongly interacting molecules, such as those attached to a surface [8] or bound to other biomolecules [16], the rotational diffusion time increases and the range of accessible rotational states is greatly diminished so that the polarization signal is more easily detected. Similarly, molecules with a higher directional structural stability such as DNA intercalating dyes [17], [18], fibers [19], or membrane proteins [20] show a strong polarization signal if care is taken not to depolarize the emitted light.

This effect can be utilised in ensemble scale fluorescence anisotropy assays, commonly used for drug discovery [21] and studies of enzymatic binding [22] and nucleic acid conformations [23]. While anisotropy assays can be refined to a dilute single-molecule level *in vitro* [9] or to high-sensitivity using strategies such as modulated input polarization [24], the fact remains that without detection of single molecular complexes in individual living cells, this approach can neither describe heterogeneous molecular phenotypes across living populations, nor disentangle the dynamic microenvironments in the vicinity of these molecules within each cell. However, polarization-sensitive imaging of individual live cells [25], [26] can be combined with single-molecule techniques [27], [28] which directly enable this level of discrimination, as previously used to remove bias from super-resolution reconstructions [29], [30]. These combined super-resolution techniques have been used to study dynamic events in macromolecule assembly with full molecular orientation reporting at a timescale of 100 ms [28], [31]–[33], although the requirement for total internal reflection renders this primarily suitable for surface-based imaging. Sophisticated theoretical treatments also exist for the unequivocal detection of molecular orientation in images [34], [35], though most require defocus or other image aberrations that are not ideal or intuitive for live imaging.

Techniques in which the excitation polarisation is modulated [36]–[40] have promising signal-to-background characteristics and potential for true 3D reconstruction of molecular orientation [40]. Polarization-based orientation information is also weakly encoded in an emitter's point-spread function [41], although this information is difficult to extract without unequivocal single-molecule emission. A major benefit of single-molecule localization microscopy (SMLM) techniques such as polar-dSTORM (where STORM denotes stochastic optical reconstruction microscopy) [27], [42] is that the pointing flexibility of the probe can be statistically decoupled from the orientation of the biological target, with exquisite precision. However, the heavy requirements for angular and spatial sampling in these approaches preclude the study of highly dynamic molecular complexes and assemblies *in vivo*, particularly within the relatively low photon budget of fluorescent protein fusions compared to bright organic dye reagents.

Away from polar-dSTORM, polarization microscopy has also been combined with structured illumination microscopy (SIM) [38], where actin fibrils were imaged giving orientation as a trivial readout of the reconstructed SIM image, and photoactivated localization microscopy (PALM) [43] where the photoactivation was used to study orientation inside a nuclear pore complex on a molecule-by-molecule basis. Polarized super-resolution microscopy is also compatible with multi-photon excitation [44], which allows for deeper imaging into optically heterogeneous samples such as biological tissues at higher spatial resolution, although the range of fluorescent fusions and capacity for dynamic tracking or time-lapse imaging are limited.

In our present work, we report an easy-to-implement method to combine super-resolvable Slimfield microscopy [45] with single-molecule polarization microscopy, in which we demonstrate as proof-of-concept can trivially image single fluorophores to sub-pixel lateral precision at moderately rapid exposure times of 40 ms, splitting the linear *s* and *p* polarization components and imaging them simultaneously onto separate halves of the same camera pixel array. The use of Slimfield microscopy means we are not limited to surface imaging, such as total internal reflection fluorescence (TIRF) microscopy and can therefore image biological structures *inside* cells when appropriate. The image stacks are analyzed with our in-house MATLAB software package called ADEMSCode [46] which identifies the positions of the fluorophores to super-resolved precision, nominally to a lateral spatial precision of ~40 nm for typical fluorescent proteins at comparable integration timescales [47]. With post-processing software written in Python, we spatially register the two polarization detection channels using sub-pixel phase cross correlation-based transformation functions and find the total integrated pixel intensity for the lateral component of each detected diffraction-limited fluorescent focus in each polarization channel, converting this into a polarization value, on a fluorescent molecule-by-molecule basis. By reconstructing the two-channel image into a single channel, we can also estimate the molecular stoichiometry of *in vivo* protein complexes by measuring the initial intensity of each fluorescent focus prior to any photobleaching and then normalizing this against the measured total integrated intensity of a single fluorophore [48], denoted here as the Isingle value, using our Python implementation of the fluorescent foci tracking and stoichiometry quantification algorithm [49].

We first demonstrate that under circularly polarized excitation, monomeric GFP (mGFP) either immobilized to a glass coverslip surface or diffusing freely in solution recovers the expected qualitative form of the polarization distribution. We then demonstrate the utility of our technique using three different live cell systems of the Mig1 glucose sensing pathway in *Saccharomyces cerevisiae* (budding yeast), the Rep accessory helicase protein in *Escherichia coli* bacteria and the aggresome stress response organelle also found in *E. coli*. Finally, we apply our method to study the pattern of fluorescence polarization of single DNA molecules that have been combed out onto a coverslip surface *in vitro* and labelled with a fluorescent dye that binds with a well-defined orientation relative to the DNA axis. Our aim here is not to quantify the exact orientation of each emitter. Instead, we get single-particle stoichiometry and diffusion information and aggregate polarization data so that we can compare populate-level polarizations and binding behavior.

The Mig1 glucose repressor in budding yeast provides a model system for regulated protein assembly in eukaryotic cells that can be controlled by the extracellular concentration of glucose;   under plentiful glucose supply, Mig1 is concentrated inside the nucleus to suppress a host of metabolic pathways concerned with the metabolism of glucose, including several implicated in diabetes and cancer in humans [50].  However, when yeast cells are incubated in low glucose conditions, the localization bias of Mig1 molecules shifts away from the nucleus into the cytoplasm leading to an increase in the mean nearest neighbour distance of fluorescent foci and a lower expectation of optical overlap of diffraction-limit fluorescent foci, despite a higher total expression of Mig1 molecules per cell [51].

Rep is a so-called "accessory" helicase present in bacteria that is associated with the molecular machinery responsible for DNA replication, an intricate mesoscale complex called the replisome, via "replicative" helicase DnaB [52] and helps in clearing nucleoprotein barriers to replication as well as restarting replication of DNA after replisome stalling due to these barriers. It has been shown previously by our group that clusters of Rep in live cells are primarily hexameric in terms of their stoichiometry [53]. We chose to look at this protein here as it has a relatively well-defined low stoichiometry and therefore serves as a good test of our correlative stoichiometry/polarization detection method. We also apply our approach to a membraneless bacterial intracellular organelle called the aggresome, which forms inside bacteria as a result of stressful environmental conditions [54]. Using our live cell single-molecule fluorescence polarization imaging we show that all three of the protein assembly systems of Mig1, Rep and the aggresome have polarization distributions like those of freely diffusing GFP. At low stoichiometry values corresponding to just a handful of molecules per protein assembly we observe a broad variance in polarization reflecting population sampling of individual polarization states, while as the stoichiometry increases to typically over ~10 molecules per complex the polarization variance narrows to closer reflect the ensemble average of the unpolarized values. The stoichiometries we measure are in line with previously reported values, and this technique therefore represents a promising tool for interrogating polarization and stoichiometry states simultaneously of complex protein assemblies in single live cells. An added feature to the fluorescent foci tracking is that the effective apparent diffusion coefficient can be also be measured, therefore our tool enables correlative synchronous measurements of molecular content, polarization, and translational mobility at the level of single tracked molecular complexes and assemblies.

For experiments on DNA *in vitro* we use a specific DNA-binding dye called YOYO-1, that binds with a fixed orientation when intercalated inside the major groove of DNA [18]. The polarizations recorded are indicative of DNA substrate orientation as expected from our knowledge of YOYO-1 binding.

**Methods**

*Slimfield Microscopy*

Slimfield microscopy was performed using a bespoke epifluorescence microscope described previously adapted for single-molecule detection with millisecond timescale sampling [55] . The excitation source was a continuous wave laser at 488 nm (Coherent Obis XS) with intrinsic vertical polarization. The settings were as follows for all experiments:  source power 20 mW equivalent to approximately 8 kW/cm$^2$ at the sample after beam expansion and de-expansion (see Figure 1); integration time per frame of 40 ms (i.e., frame rate of 25 Hz).  For experiments requiring circularly polarized excitation, a λ/4 waveplate (Thorlabs part AQWP05M-600) was placed in the laser delivery path prior to the microscope entrance pupil. The correct circularization was ensured by rotating the λ/4 waveplate to equalize the power transmitted through a linear polarizer in the sample plane, independently of the linear polarizer's direction. For experiments using linearly polarized excitation, this λ/4 waveplate was removed and replaced with a λ/2 wave plate (Thorlabs part AHWP05M-600) rotated by either 0° for vertical polarization or 45° for horizontal polarization at sample. For all microscopy we used a Photometrics Prime 95B camera and a Nikon Plan Apochromat (NA 1.49) objective lens.

*Orthogonal polarization signals detection*

The polarization splitter module centred on a ratio polarizing beam splitter cube (Thorlabs PBS251), into whose downstream paths linear polarizing filters (Thorlabs LPVISA100) were placed to clean up traces of non-polarized light. The polarization splitter module was fixed to a magnetically mounted breadboard to allow the exchange of this module for the color channel splitter usually in place. A schematic diagram of the polarization splitter can be seen in Figure 1.

*Purified GFP in vitro sample preparation*

Monomeric green fluorescent protein (normal enhanced GFP but with the addition of an A206K point mutation to inhibit dimerization) was purified from competent *E. coli* as described previously [47]. Samples were prepared inside 'tunnel' slides with standard microscope slides, plasma cleaned #1.5 coverslips and double-sided sticky tape as described previously [48], [53]. Initially, the tunnel was washed with 100 μL of phosphate buffer saline (PBS, Sigma Aldrich), and then 10 μL of 10 μg/mL anti-GFP (RPCA-GFP, EnCor Biotechnology Inc.) was flowed in and incubated for 5 min. The antibodies form a sparse carpet on the plasma-cleaned coverslip surface for mGFP to bind to. Excess anti-GFP remaining in solution was then washed out with 100 μL PBS, and the slide was then incubated with 10 μL 1 mg/mL bovine serum albumin (BSA) (05479, Sigma-Aldrich) for 5 min to passivate the surface against non-specific binding of mGFP. After a further 100 μL PBS wash, 10 μL 50 ng/mL (2 nM) purified mGFP in PBS was flowed in and incubated for 5 min, to bind to the surface immobilized anti-GFP. A final 100 μL PBS wash was performed, the slide sealed with clear topcoat nail polish (Rimmel London) and imaged directly afterwards in Slimfield. For mGFP in solution, the same procedure was used with the omission of the anti-GFP incubation step, and the final incubation/wash steps, and focusing a few microns into solution as opposed to on the coverslip surface itself as for surface immobilized mGFP. In this case, the BSA incubation passivates the entire surface and the mGFP, therefore, remain in solution for imaging of their diffusive behavior. All incubations were performed with the slide inverted in a humidity chamber at ambient room temperature. A schematic diagram of the immobilized mGFP assay is shown in Supplementary Figure S1.

*DNA/YOYO-1 sample preparation*

YOYO-1 was diluted in PBS to 1 µg/mL and mixed with a 1:9 dilution of lambda DNA:PBS in equal volumes as described previously [46]. To prepare the coverslip surface, coverslips were plasma cleaned for one minute in atmospheric plasma, and the clean side was then coated in 0.1% (w/v) poly-L-lysine and incubated for two minutes. The coverslip was then washed in MilliQ water and allowed to partially air dry so that no large droplets of water remained, but the surface was moist. To assemble the microscopy sample, 5 µL of the DNA/YOYO-1 mixture were pipetted onto a glass microscopy slide. The coverslip was then dropped onto the sample droplet with the poly-L-lysine-functionalized side in contact with the DNA/YOYO-1 droplet to facilitate surface immobilization, resulting in a molecular combing effect to stretch out single DNA molecules onto the coverslip surface [46]. The coverslip edges were sealed with nail polish and the sample was imaged immediately as described above.

*S. cerevisiae Mig1-GFP sample preparation*

*S. cerevisiae* of either BY4741 parent strain, or genomically labelled at the glucose regulator, Mig1-GFP, were grown on Yeast Extract–Peptone–Dextrose (YPD) plates + 4% glucose for 48h at 30°C. The cultures were inoculated into liquid Yeast Nutrient Broth (YNB) media with 4% glucose with sequential dilutions and grown overnight, then transferred either into YNB + 4% (high glucose condition) or +0.2% (low glucose condition) for 1h before imaging as described above.

*E. coli Rep-mGFP sample preparation*

A previously described *E. coli* Rep-GFP labelled cell strain [53] was grown in LB broth overnight at 37°C to saturation. The overnight liquid cell cultures were then diluted 1000-fold in 1 × 56 salts minimal medium fortified with 0.2% glucose and grown to mid-log phase at 30°C. To prepare slides for microscopy, the cells were spotted onto a 1% agarose pad containing 1 × 56 salts and 0.2% glucose and affixed to a glass microscope slide. Finally, a glass coverslip was laid over the top and cells were imaged immediately as described above.

*E. coli aggresome sample preparation*

To induce aggresome formation, *E coli* cells cultured overnight were diluted by 1:1000 into fresh LB medium and grown at 37 °C for 24 hours to induce nutrient stress [54]. To prepare slides for imaging, the cell washed two times with M9 minimal medium and spotted onto a 1% agarose pad containing M9 medium and affixed to a glass microscope slide. Then a glass coverslip was laid over the top and the sample was imaged immediately as described above.

*Choice of polarization metric*

The fluorescence polarization metric, $\rho$, is relatively intuitive and is well-defined within the range (-1, +1) for single detected fluorescent foci (the diffraction-limited point spread function in the microscope's focal plane).

$$\rho_V = \frac{I_{VS} - I_{VP}}{I_{VS} + I_{VP}}$$

$$\rho_H = \frac{I_{HS} - I_{HP}}{I_{HS} + I_{HP}}$$

$$\rho_C = \frac{I_{CS} - I_{CP}}{I_{CS} + I_{CP}} \approx \frac{\rho_V + \rho_H}{2}$$

Both the method presented here, and standard anisotropy assays aim to measure the same fundamental property of fluorescence polarization, and so share the assumption that fluorophores are dipolar and act independently. However, there are non-trivial differences, including their assumptions about the number of emitters per measurement, or equivalently the stoichiometry $S$ of a detected fluorescent focus. Most of these assays describe ensemble measurements (*Stoichiometry* >> 1) of anisotropy, $r = 2\rho_V/(3 - \rho_V)$, with perpendicular axes of excitation and detection. Under Slimfield microscopy, the single- or few-emitter limit is important - within which the relevance of the anisotropy metric is unclear - and the optical axes of excitation and detection are colinear. With care, one may assess the theory used for routine assays, assuming the collective behavior of randomly oriented ensembles, to interpret the Slimfield polarization results. For example, the Perrin equation describes how under linear polarized excitation, the photoselected polarization signal decreases according to the rotation timescale $\tau_R$, which is the property of interest in our experiments. It reads as:

$$\left(\frac{1}{\langle\rho_L\rangle} - C\right) = \left(\frac{1}{\rho_L'} - C\right)\left(1 + \frac{\tau_F}{\tau_R}\right)$$

where $\tau_F$ is the fluorescent lifetime and $C = 1/3$ in the ensemble limit ($S >> 1$) and the subscript $L$ can refer to either $V$ or $H$. The *fundamental polarization, $\rho'$*, describes the theoretical, integrated response of many randomly oriented emitters of in the *absence* of rotational depolarization. The concept can be extended to single or few emitters, for which the expectation in general depends on the stoichiometry $S$, i.e., the number of independent emitters per focus. Under excitation that is parallel to one of the detection channels, the photoselection effect favors that channel and the expectations become $\rho_V' = (1 - 1/\sqrt{S})/2$ and $\rho_H' = -(1 - 1/\sqrt{S})/2$ respectively [56].

Under circularly polarized or unpolarized excitation, the situation appears rather different. The photoselection effect is equalized over both axes of the detector and the resulting expectation is unbiased, $\rho_C' = (\rho_V' + \rho_H')/2 = 0$. However, the rotational decay of each measured non-zero polarization is inherently the same. Since anisotropy $r$, and not polarization $\rho$, is additive in the ensemble limit [21], the apparent rotational timescale is related to the average of $(1 + \tau_F/\tau_R)^{-1}$ over the measurement time $\tau$, which in this case is the camera exposure time of 40 ms.

As such, the polarization signal from an individual fluorescent focus provides a measure of the *fastest* rotational timescale of the emitters at that location. In the context of our experiment therefore, a focus with non-zero polarization signal indicates a set of molecules, within a super-resolvable localization, that are persistently constrained in their rotational dynamics over the full duration of the exposure.

In all cases, the measured polarization signal from our microscope instrumentation if applied to cell samples is also attenuated due to several depolarization factors. With colinear detection from a monolayer of cells, the contribution due to scattering is minimal. The largest contribution is expected to be the high numerical aperture (NA) of the objective, which distorts how polarization components in the sample plane couple to those in the excitation and detection paths. We note that in Slimfield, the excitation beam strongly underfills the back focal plane of the objective lens to emerge collimated, and therefore the incident laser itself will not be prone to depolarization from the lens' numerical aperture. However, the theoretical effect on depolarization of the fluorescence emission [57] reveals a similar form to the Perrin equation above:

$$\left(\frac{1}{\rho} - 1\right) = \left(\frac{1}{\rho_0} - 1\right)\left(\frac{2}{1 + \sin(2\psi)/2\psi}\right)$$

where $\rho_0$ is the underlying polarization in the limit NA = 0 and $\psi = \sin^{-1}(\mathrm{NA}/n)$ is the half angle of the detection, with *n* the refractive index of the objective's coupling medium. The effect is such that extreme polarizations remain accurate, but smaller polarization signals are suppressed by up to 40% at our NA=1.49.

A locally-variable contribution to depolarization is homo-FRET [58], in which photoselected emitters transfer their energy to another nearby emitter, whose alignment has a weaker correlation with the absorbed photon. The net effect is to depolarize the emission at that location. As such, homo-FRET itself has been identified as a potential signature of protein aggregation, but it only occurs when the chromophores approach within a few nanometers, so the corresponding depolarization is a weak effect for relatively large fluorescent proteins.

For the work presented here, the above effects limit the ability to infer a quantitative molecular orientation, which would require a high degree of confidence in the corrections for depolarization. Instead, we look at population-scale measurements by summing the responses of individual fluorescent foci, themselves not inherently of single fluorophores if applied to live cell samples, to allow a qualitative interpretation of the polarization distributions.

*Image analysis*

Images were analyzed with ADEMSCode [46], a home-written package in MATLAB (MathWorks), to identify candidate foci corresponding to fluorescent complexes. The super-resolved positions of the foci with an integrated signal-to-noise ratio (SNR, equivalent to the amplitude of fitted 2D Gaussian function divided by the standard deviation of the local background pixel intensity noise multiplied by the area of the putative fluorescent focus) of at least 0.4 for mGFP assays was taken and used for customized Python post-processing. Specifically, the full frame was split into two rectangular regions of interest and the translation-only mapping between them was found with scikit-image's *phase_cross_correlation* function [59]. Not only was this mapping used to transform the images of the second channel onto the first, but also to shift the channel 2 spots into their locations in channel 1. The integrated intensity of each fluorescent focus in each channel was found by summing the intensity inside a circular mask of radius 5 pixels centered on the super-resolved position after local background correction. The local background was calculated as the mean average of pixels specified by a bitwise XOR between the circular spot mask and a square of side length 17 pixels also centered on the fluorescent focus locus. The corresponding values in both channels for a given fluorescent focus were used to calculate the polarization

$\rho = (I_1 - I_2)/(I_1 + I_2).$

Together with the masks, these values were also used to plot polarization heatmaps. To avoid double counting of fluorescent foci visible in both channels, any candidates in channel 2 closer than 2 pixels to any candidate in channel 1 were neglected. All plots were made with *matplotlib* [60]. The schematic of this process is shown in Supplementary Figure S2.

We found that the emission polarization distribution appeared to vary spatially in a similar manner to the intensity of the incident laser, whose beam profile underfills the field of view. Specifically, we found that outside the central illuminated region the polarization was skewed positive, while in the center the skew appears negative. The outer region is not illuminated by the laser and therefore must be an artifact not corresponding solely to the ratio of emitted fluorescence. The negative skew in the center cannot similarly be dismissed, although the background in channel 2 (vertically polarized) is everywhere significantly larger than channel 1 (horizontally polarized) (Figure 2 a,b) which suggests a negative bias in the polarization signal that may not be fully compensated by our existing method of background subtraction.

We restricted our downstream analysis to the fluorescent foci located inside a circle of radius equal to the full width at half maximum of the beam, within which the excitation intensity (and the expected total emission intensity) is relatively high. In practice this radius is approximately 90 pixels or 4.8 µm in the sample plane.

For the DNA/YOYO-1 images, the full channels underwent registration, and the polarization $P$ was calculated on a pixel-by-pixel basis to create a heatmap of polarization. However, the raw image intensities were prone to large background which effectively washed out the negative polarization signal. To compensate for this, the channels were scaled such that the maximum intensity in each channel was equal. Although this precludes quantitative analysis of the generated polarization heatmap, it nonetheless demonstrates the presence of the distinct polarization states associated with molecular orientation.

*Calculating the brightness of single dye molecules, Isingle*

The *E. coli* aggresome data was imaged until it was fully bleached and in the photoblinking regime and was reconstructed by registering both channels on to each other and summing. This reconstructed single-channel image was then analyzed with our new Python single-molecule tracking code PySTACHIO [49] which plots the integrated foci intensity and finds the peak of a kernel density estimation fit to the intensity distribution. We also checked this against the surface-immobilized mGFP data and both were found to give a consistent Isingle value around 130-140 integrated pixel values, equivalent to a quotient of 70 ± 8 (mean ± s.d.) photoelectrons frame$^{-1}$ molecule$^{-1}$, which is consistent with a shot-noise limited measurement.

**Results**

*Vertically, horizontally, and circularly polarized light give different distributions for mGFP immobilized in vitro*

We began by immobilizing mGFP as in the protocol in Figure 2, and imaging with the excitation beam polarized either vertically, horizontally, or circularly. We acquired >10 fields of view in each case and analyzed as above. In Figure 4 we present representative fields of view under circularly polarized excitation (Figure 2 a,b) and extracted a polarization heatmap (Figure 2c). It is possible to resolve the apparent net polarizations of individual mGFP molecules at this imaging speed despite the large uncertainty (up to 50%) in their total emission intensity. We see that in the cases of vertically and horizontally polarized excitation (Figure 4 panels d and e respectively) there are distinct distributions (Kolmogorov-Smirnov (KS) test [61], $p < 0.01$) which are skewed towards the polarization of the excitation laser as expected (positive for horizontal and negative for vertical, Table 1).

|             | $H$   | $V$   | $C$  | $\frac{H+V}{2}$ | $C$ (fd.) |
|-------------|-------|-------|------|-----------------|-----------|
| $H$         |       |       |      |                 |           |
| $V$         | 0.008 |       |      |                 |           |
| $C$         | 0.7   | 0.02  |      |                 |           |
| $\frac{H+V}{2}$ | 0.06 | 0.8 | 0.1  |                 |           |
| $C$ (fd.)   | 0.1   | 0.003 | 0.2  | 0.03            |           |

**Table 1:** Two-sample Kolmogorov–Smirnov test (MATLAB *kstest*) of dissimilarity of polarization distributions, showing $p$ significance values. Labels refer to the following datasets: immobile mGFP imaged with horizontally ($H$), vertically ($V$) or circularly ($C$) polarized excitation, or freely diffusing mGFP imaged with circularly polarized excitation ($C$, fd.). Very low values $p < 0.1$ (black) imply a significant difference such as a shift in median, while moderate values around p = 0.1 (grey) may result from a similar median but different variance, or *vice versa*. High values, p >> 0.1, indicate that the test cannot separate the distributions. The $C$ distributions appear more similar to $H$ than to $V$, due to the negative polarization bias in measurement.

Physically, this arises because of photoselection, whereby fluorophores aligned parallel to the polarization of the excitation laser are more likely to be excited than those aligned perpendicularly. This leads to a higher rate of detection of aligned fluorophores and the distribution overall is therefore skewed towards the excitation polarization. The magnitude of photoselection bias here is expected to be about $\rho$ = ±0.3, qualitatively consistent with observation. Quantitatively, however, we cannot exclude the presence of confounding factors of a similar magnitude. Some are expected to average out in the distribution, such as the noise on each single-molecule polarization measurement, while others including depolarization and $G$-type correction factors will not.

Under circularly polarized excitation, symmetry considerations would suggest a distribution which is the sum of the vertical and horizontal cases, and indeed we see in Figure 2f that the circularly polarized distribution is qualitatively similar to the sum of the distributions in panels 2d and 2e (KS test, $p$ = 0.1). This acts as a useful check on the delivery of excitation and of the consistency of detection. The shape of this distribution resolved at a high statistical power is also reassuringly symmetric around its mean, since the photoselection is equalized along both axes. However, rather than the expected mean $\langle \rho_C \rangle$ = 0 for the circular polarization case, there is a consistent negative offset, which strongly indicates a significant difference, of order 30%, in the optical transmission efficiencies, and/or depolarization properties, of our split detection channels.

However, we are not seeking to extract more detailed orientation information for individual dipoles and noting the differences in overall distribution for bound and free fluorophores, we can say that this source of systematic error, similar to a $G$ correction factor in anisotropy instruments, does not materially affect the qualitative interpretation of our results. Future calibrations may remove these influences such that the polarization signals can be rendered independent of the instrument.

The relative proportion of fluorescence intensity in either polarization detection channel across all the surface-immobilized mGFP assays we tried varied between approximately 1% and 99% as a proportion of the sum of $I_1+I_2$.

*mGFP freely diffusing shows a distinct polarization distribution*

In Figure 2g we show the overall distribution for tracking mGFP molecules freely diffusing *in vitro*. In total we tracked 10 different acquisitions for 100 frames each giving 1,000 total frames of information in this case. The polarization distribution is smooth, symmetric, and again centered around approximately $\rho = -0.2$, which is distinct from the immobilized mGFP cases under linear excitation (Figs 2d-e, KS test: $p < 0.1$), but with far fewer extreme values when compared to the immobilized circular excitation (Fig 4f) or the sum of the immobilized linear excitation cases (KS test, $p < 0.1$). The expectation for a freely diffusing system would be that the polarization distribution peaks around $\rho = 0$ regardless of excitation polarization, as the intensity in each channel should be approximately equal under rotation events during the fluorescence lifetime (which washes out any photoselection under linear excitation) and under many thousands of rotation events during the camera integration time (which mask the presence of a dipole under circular excitation).

The negative offset manifests in a noticeable shift of the mean polarization, though the decay is symmetrical on both sides of the center of the measured distribution. This is the expected behavior for a system with a consistent sample-independent bias in polarization measurement, likely due to rectifiable differences in the noise floors and optical properties of the two channels. Regardless of this systematic error, there is a clear similarity (KS test, $p > 0.1$) in the averages of the immobile and freely diffusing cases under circular excitation, while the tails of the distributions are qualitatively distinct. The apparently narrower distribution in the freely diffusing case would imply a more rotationally averaged dipole as expected, but this difference in variance cannot presently be separated from the contributions due to the underlying sensitivity of the measurement.

*YOYO-1 intercalated DNA shows a polarization signal dependent on its orientation*

In Figure 3 we show the results of imaging DNA surface immobilized on the cover slip with YOYO-1 intercalated. YOYO-1 is known to be strongly bound to DNA with a fixed orientation making it a useful reporter dye for single-molecule polarization [18]. Here we see qualitatively that the vertically oriented DNA has a positive (i.e., horizontal polarization) signature - because the YOYO-1 is perpendicular to the helical axis of the DNA. The polarization microscopy is able to discern these orientations though as we have a one-shot imaging system, we are not able to accurately determine the precise orientation of a dye. This is however evidence that the orientational polarization response is captured correctly by our imaging and analysis methodology.

*Rep-GFP in E. coli shows a similar distribution of polarization to freely diffusing mGFP*

Figure 4 shows the results of our imaging Rep-GFP in living *E. coli*. Most strikingly the 2D histogram in Figure 4a shows the convergence of polarization to our mean free-diffusion value as stoichiometry increases. This indicates a random averaging of orientations with respect to the detector as would be expected for large stoichiometries, as the increased number of uncorrelated fluorophores has the effect of scrambling the average polarization signal and therefore giving the appearance of neutral polarization. However, the measurable spread of polarization values at low stoichiometry indicates that for small complexes polarization signals can be extracted that are consistent with alignment between the electric dipole axis of the GFP fluorophore and the electric field of the excitation laser.

The stoichiometry distribution (Figure 4b) is in line with previously estimates of Rep that is not colocalized with the DNA replication fork, known to have a broadly hexameric trend though less pronounced than that for Rep colocalized with the replication fork [53]. Indeed, as the majority of Rep molecules expressed in any given the cell are not likely to be colocalized with the replisome in a given sampling time window, we expect our polarization measurement to be representative of this fraction.

*The E. coli aggresome shows a null polarization distribution*

In Figure 5 the stoichiometries and polarizations of *E. coli* aggresome foci are reported. Again, in Figure 5a we see that the polarization converges on the ensemble mean value for the freely diffusing case as stoichiometry increases. In the case of aggresomes, we need to consider higher stoichiometries than other samples, since they are likely a compact agglomeration of several different proteins in a small, confined area, thus several molecules can potentially arrange randomly in generating an ensemble polarization signal over the whole aggresome. Here, we see a tight peak around the null polarization value, which suggests three possibilities - one is that the aggresome itself is rotating rapidly and giving a 0 polarization signal; second it could be that the proteins and/or their fluorescent tags within the aggresome are free to rapidly rotationally diffuse; or thirdly it could be that the aggresome is made up of tightly packed proteins and is relatively static, but proteins of different orientations combine to scramble the polarization signal to 0. Given previous work on aggresomes, we believe the latter to be more likely to be the case. However, further work, such as using photoactivated dyes, would be needed to build up a map of individual protein polarizations, but we present the work here more as proof-of-concept to show the potential of our new tool. In Figure 5b we should the 2D histogram of diffusion coefficients and polarizations which shows no obvious trend at these scales. Once again, the polarization distribution shape (Figure 5c) is qualitatively similar to that of the freely diffusing mGFP, though marginally narrower. The stoichiometry (Figure 5d) is consistent with previously reported values for the *E. coli* aggresome. In Figure 5d we demonstrate the correlative aspect of our method by plotting all quantified properties of each tracked fluorescent focus against each other in a 3D scatter. For each such fluorescent focus we are able to simultaneously measure the diffusion coefficient, stoichiometry, and polarization.

*Mig1-GFP in high and low glucose conditions*

We find that Mig1-GFP is primarily localized to the nucleus or is present predominantly throughout the cytoplasm of the cell, for high and low glucose conditions respectively as reported in previous work (Figure 6). In both high and low glucose conditions the stoichiometry values for fluorescent foci and the total integrated protein copy per cell are also consistent with previous work to within experimental error. The tracked complexes showed an average trend towards a neutral polarization distribution at all stoichiometries, but with a decreasing variance as stoichiometry increases. The trend is such that the spread decreases sub-linearly with increasing stoichiometry, which is anticipated from the deviation $\sim 1/\sqrt{S}$ expected under averaging of independent normally distributed polarization signals from randomly oriented fluorophores.

**Discussion and Conclusions**

Previous studies have reported a range of valuable instrumentation that can perform simultaneous super-resolved localization microscopy and polarization imaging, for example using structured illumination [38], PALM [43] and polar-dSTORM [42]. While each of these has distinct advantages - straightforward orientation reconstruction with SIM, molecule-by-molecule investigation of multi-protein structures with PALM, and high spatial resolution for polar-dSTORM, they have the disadvantages that go with their respective coupled technique. For example, lower effective spatial resolution in SIM, and no time-resolution as such for polar-dSTORM which used fixed cell samples. In this proof-of-concept work here we have extended rapid and high sensitivity Slimfield microscopy to image fluorophores that are commonly employed as single-molecule fluorescent protein reporters in cellular processes, using single-molecule super-resolved localization microscopy combined with simultaneous polarization information. The lateral spatial precision and integration time for imaging we report is comparable to polarization PALM tracking, however our principal innovation is to correlate polarization measurements of individual tracked fluorescent foci with their measured molecular stoichiometry in terms of the number of fluorescently labeled molecules present in each tracked particle. Note also that although we have focused primarily on reporting the polarization and stoichiometry values of tracked molecular complexes, complete measurements of the apparent diffusion coefficient may be obtained for each tracked fluorescent focus as we show in Figure 5, so our new tool has the capability to correlate polarization, stoichiometry and translational mobility at the level of single dynamic molecular complexes in live cells. We have demonstrated the core utility of this correlative approach for a commonly used fluorescent protein reporter *in vitro*, as well as three different biological systems in both live single budding yeast and *E. coli* cells.

We used our in-house fluorescent foci tracking suite of software ADEMSCode and a novel Python analysis code to spatially register image channels, detect fluorophores, and measure distributions automatically of the polarization metric

$\rho = (I_1 - I_2)/(I_1 + I_2)$, where $I_1$ and $I_2$ are the respective horizontally and vertically polarized components of the fluorescent emission.

For surface-immobilized fluorophores we show three distinct distributions depending on excitation polarization and demonstrate that the circularly polarized excitation gives rise to a distribution that is approximately the sum of the vertical and horizontal excitation distributions as expected. In the case of freely diffusing fluorescent protein, we used circularly polarized excitation light to demonstrate that the distribution is approximately symmetrical around a small negative polarization value, which we believe is an artifact due to different noise floors between the two detection channels, in contrast to previous studies in which the polarization is split by a single prism [42]. The freely diffusing case is distinct from the surface-immobilized one, indicating that even at 40 ms integration times, several orders of magnitude slower than the ~nanosecond rotational timescales anticipated in free solution, our instrument is sensitive to differences in fluorophore mobility dynamics. This is further evidenced by the clear polarization signal in YOYO-1 intercalated DNA for which the orientations of the dye molecules are well known. Although these spatially extended filamentous structures cannot be quantitatively analyzed in the same way as the single-particle tracking data, it is clear that polarization changes due to orientation can be captured by our setup, and by moving to a two-shot imaging methodology with switchable 45º polarization rotation, a true azimuthal, dipolar orientation can be captured.

In live cells, we found consistently that large protein assemblies show similar behavior to the freely diffusion mGFP case. This indicates one of two things – either within the assembly itself the molecules are free to rotate on a timescale below 40 ms, or the whole aggregate is free to rotate on that timescale. One further complication is that the fluorescent protein reporters used here are designed to be attached via a flexible linker of typically a few nm in contour length. These linkers are explicitly designed to allow some level of mobility of the reporter molecule relative to the native protein molecule itself for the purposes of helping to minimize functional impairment due to steric hindrance effects from the relatively large fluorescent protein molecule that is often comparable in effective diameter to the native protein. This linker mobility unsurprisingly may increase the likelihood that rotational diffusion of the fluorophore dipole axis is at a timescale much smaller than the smallest integration time available to Slimfield microscopy of ~milliseconds, thereby limiting the sensitivity of measurable polarization dependence. However, in the case of tightly packed molecular assemblies in live cells this effect of flexible linker mobility may be dramatically reduced, and so some polarization dependence may still be detectable.

Discerning the difference between molecular mobility inside protein assemblies and the motion of the whole assembly itself is made more complex by the polarization "scrambling" effected by the large number of emitters in a small area. As such, the variance in polarization is itself a potential predictive metric of either stoichiometry or rotational properties, if the other is known. Further work will be needed to robustly interrogate this. In particular, our module is suitable to be used also for photoactivated fluorescence microscopy and tagging molecules of interested with photoactivated dyes such as PAmCherry would allow us to see "inside" the protein aggregates we have seen in this work and to assess the orientational properties of the individual constituents independently. While *in vitro* studies have used the photoselective dependence of linearly excited polarization to infer stoichiometry [56], our approach has the great advantage that we are able to quantify both polarization and stoichiometry in each protein assembly independently and *in vivo*, which to our knowledge has not been reported previously, as well as 2D mobility information embodied in the apparent diffusion coefficient.

With brighter artificial dyes, or improved sensitivity such as through modelling of depolarization effects, we may be able to go to lower exposure times and gain more information on the fluorophores such as diffusion coefficients, and aim to use this methodology on live cells, though this is beyond the scope of the present proof-of-concept work. Although the range of power in either polarization detection channel is 1-99% the minimum intensity of the *brightest* focus from either $I_1$ or $I_2$ never goes below 50%. Since the analysis software uses the brightest detected focus from either channel to pinpoint the location of the fluorescent emitter this is what ultimately determines the lateral spatial precision. In our imaging regime, the lateral precision scales approximately as the reciprocal of the square root of the number of photons. From the number of photons detected per fluorophore relative to conventional Slimfield microscopy, we estimate that the lateral precision is better than ~60 nm.

Finally, we note that while this assay gives high spatial (super-resolved over a field of view of length scale of several tens of microns) and competitive temporal resolution (tens of ms), the polarization information is an aggregate property over an imaged population. Rather than quantify individual fluorophore orientation, we instead look at total polarization distribution to semi-quantify the overall binding behavior of the sample. Presently, it does not provide the level of sensitivity to anisotropy available in ensemble techniques, though this has a significant scope for improvement. Most notably, we use a relatively coarse analysis which only corrects for local background in each channel and does not yet fully represent the potential information contained in the images. We do not here correct for depolarization effects either in the excitation or imaging paths. In future we aim to perform more extensive and rigorously controlled calibration such that we can approximate a correction for the polarization measurement (including, but not limited to, accommodations equivalent to the instrument's *G* correction factor used in ensemble assays). Based on the speed, scale and sensitivity of our imaging method, there is future potential to extract time-resolved orientations for single molecules tracked with non-specialist, extensible, super-resolved Slimfield microscopy and to build on the information we are already able to extract such as stoichiometry. Our technique represents a first step towards developing a useful and simple to implement tool for probing the dynamical properties of molecules *in vivo* and a new avenue for understanding the physics that underlies life.

**Funding** Work was supported by the Leverhulme Trust (grant RPG-2019-156) and the Engineering and Physical Sciences Research Council EPSRC (grant EP/T002166/1).

**Data and software availability** ADEMSCode is available at https://github.com/awollman/single-molecule-tools and PySTACHIO is available at https://github.com/ejh516/pystachio-smt. All data available upon request.

**Figures and captions**

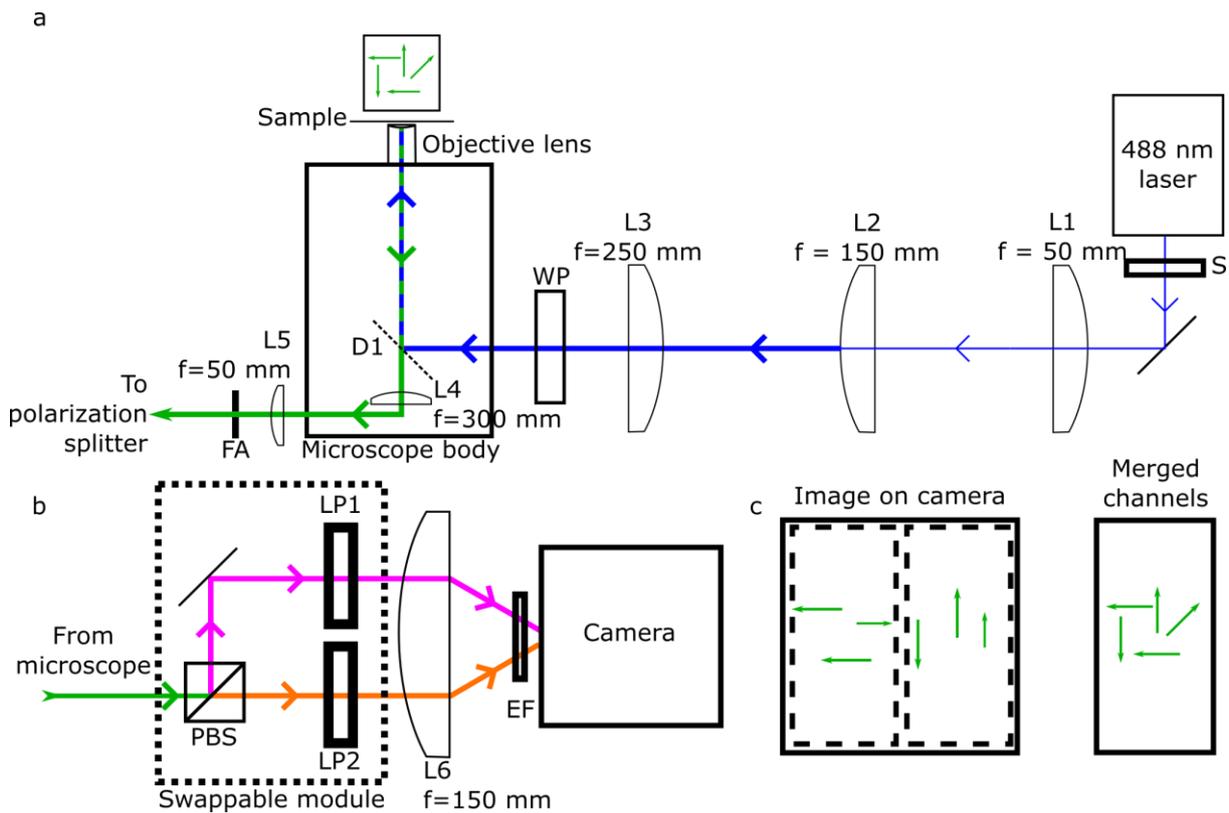

Figure 1. Schematic diagram of the Slimfield microscope. *a*) Laser light vertically polarized at source (blue) passes through a shutter (S) and is expanded 3x by the lens pair L1 and L2, before passing through L3 which forms a telescope with the objective lens to ensure light at the sample is collimated. Finally, in the excitation path the laser light passes through a wave plate (WP) on a rotational mount, either λ/2 for linear polarization or λ/4 for circular polarization. The light is directed to the sample with a dichroic mirror which allows the collected fluorescence (green) to pass through it. The emission then encounters the lens L4 and is focused at the side port of the microscope body to form an intermediate image plane into which we place adjustable slits to provide a rectangular field aperture (FA). The emission is then recollimated with the lens L5; *b*) The image light then encounters the polarization splitting module and the vertical and horizontal polarized light (orange and pink respectively) are separated by a broadband, polarizing beam splitter cube. Each polarization channel then is purified by a linear polarization filter (LP1 and LP2) before being focused on to the left and right sides of the same camera chip by the lens L6. For convenience, components inside the dotted box are mounted to a breadboard which is on removable magnetic mounts and can therefore be easily swapped for another module, e.g., color splitter. Immediately

before the camera, reflected excitation light is removed by an emission filter, EF. c) The left-hand side of the acquired image shows the horizontal polarized light, and the right-hand side shows the vertical (individual channels are indicated by dashed boxes). By registering the image and creating a composite image we recover the true fluorophore distribution.

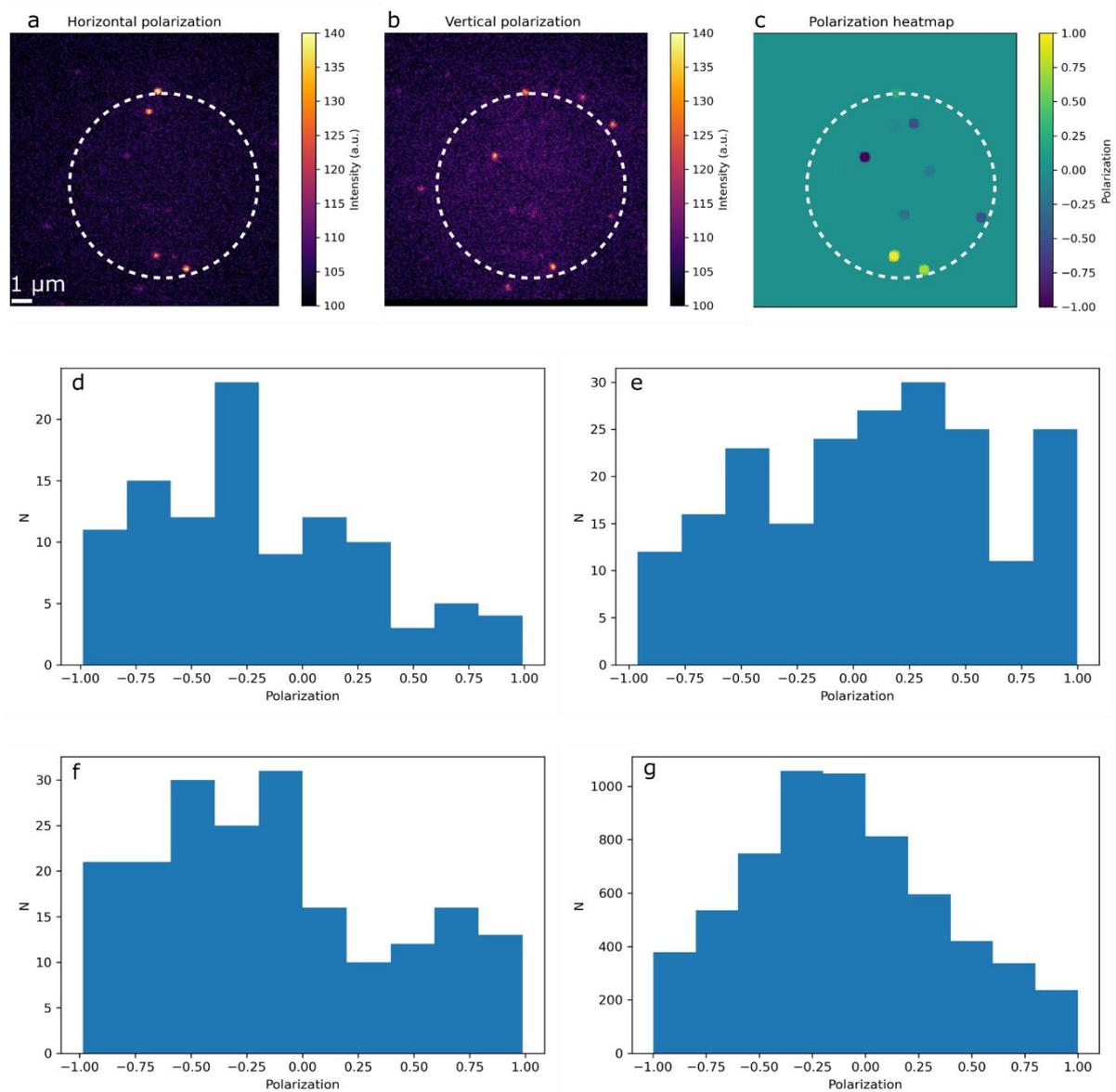

Figure 2. a) Vertical polarization channel from a representative GFP *in vitro* assay under circular polarized excitation. b) Horizontal polarization channel from the same assay as in a), both with illuminated region of interest shown as dashed circles. c) The polarization heatmap found by combining a) and b) and analyzing as in Figure 3. d-f) Polarization distribution for surface

immobilized mGFP when illuminated by d) vertically, e) horizontally and f) circularly polarized light. g) Polarization distribution for freely diffusing mGFP illuminated with circularly polarized light. Panels d-f have N = 104, 208, 195, and 6,170 tracks respectively.

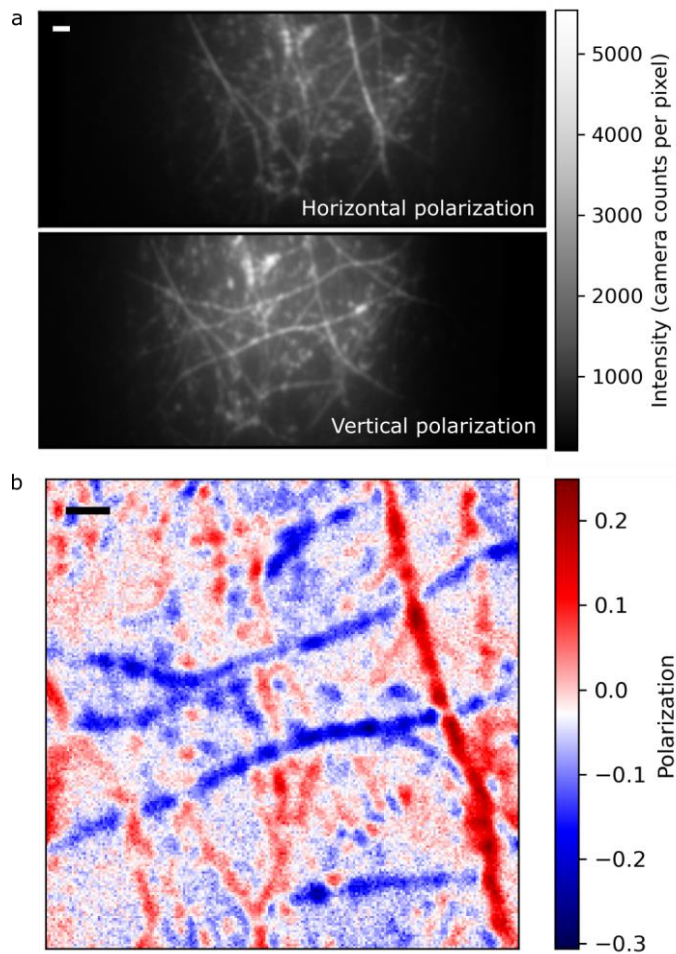

Figure 3: a) Horizontal and vertical polarization fields of view for DNA/YOYO-1 complexes (top and bottom, respectively). The horizontal DNA is more visible in the vertical polarization channel because the YOYO-1 intercalates perpendicular to the DNA helical axis; b) Heatmap of polarization demonstrating the qualitative polarization difference between differently oriented molecules. Bars: 1 µm

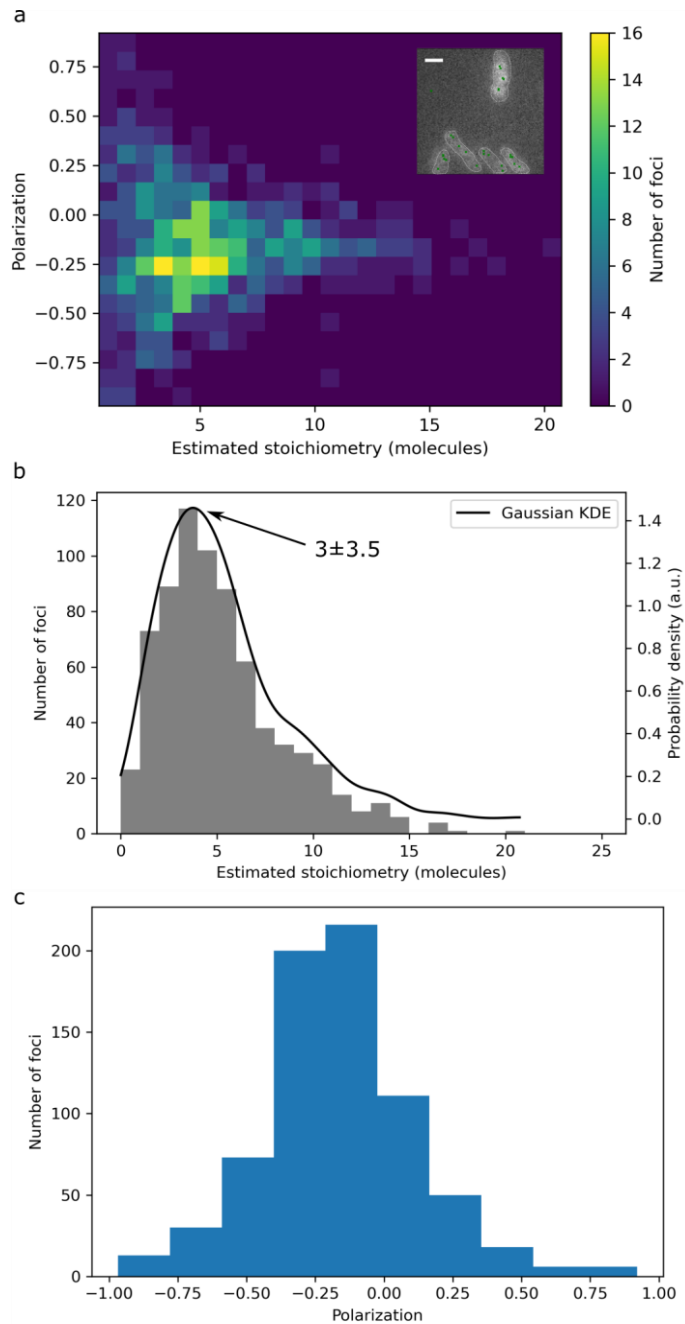

Figure 4: a) 2D histogram showing polarization as a function of stoichiometry. Inset: a representative *E. coli* field of view with tracked loci overlaid with green crosses and cell boundaries marked with white dotted lines; b) stoichiometry distribution for Rep-GFP foci with a peak at 3±3.5 molecules (mean ± half width at half maximum); c) polarization distribution for Rep-GFP foci showing qualitatively similar behavior to freely diffusing mGFP (Figure 2). Panels a bar: 1 μm

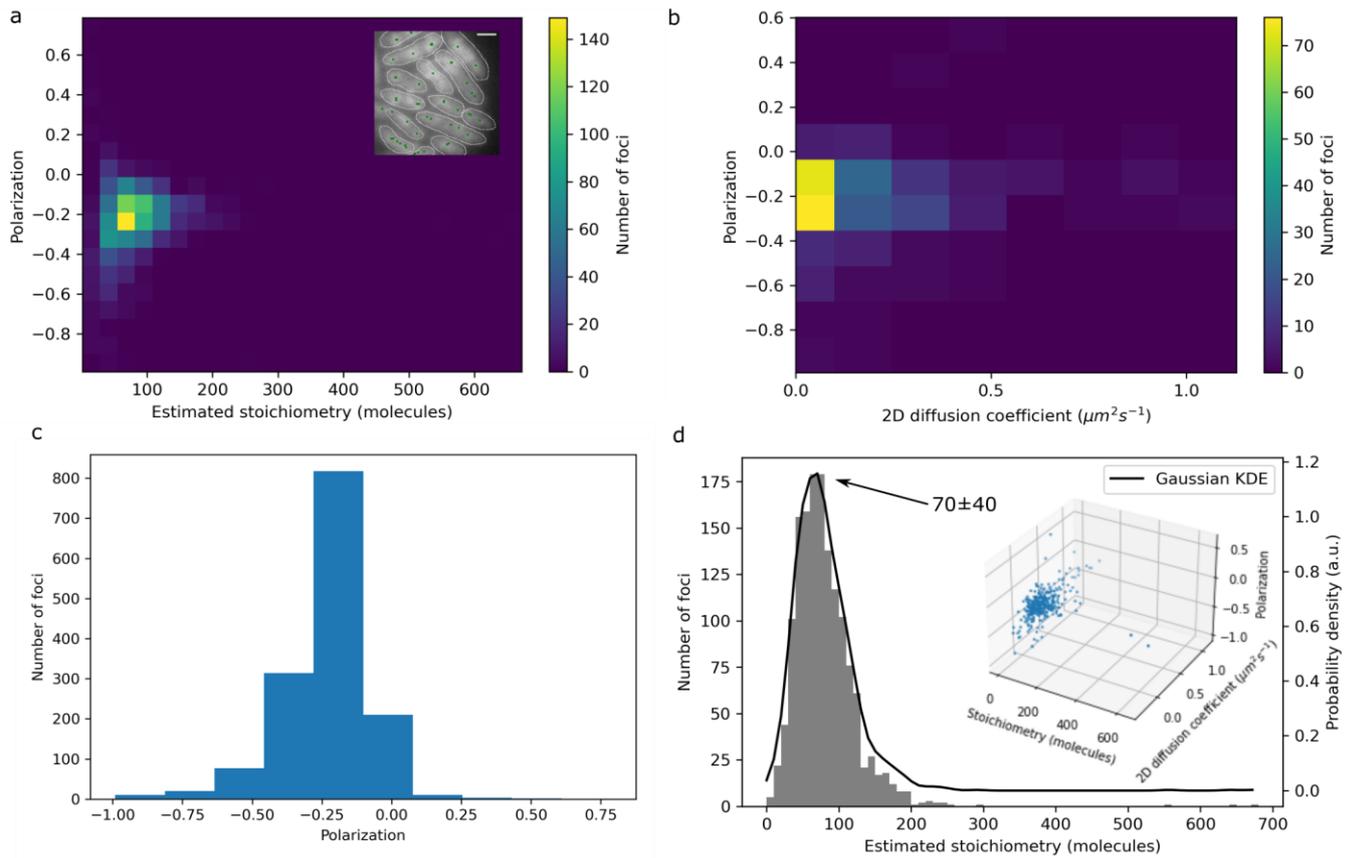

Figure 5: a) 2D histogram showing *e. coli* aggresome polarization as a function of molecular stoichiometry. Inset: a representative field of view with tracked aggresome foci overlaid as green crosses and approximate cell boundaries shown as white dotted lines; b) 2D histogram of 2D diffusion coefficient against polarization; c) polarization distribution for aggresome foci showing qualitatively similar behavior to freely diffusing mGFP (Figure 2); d) stoichiometry distribution for *E. coli* aggresome foci with a peak at 70±40 molecules (mean ± half width at half maximum). Inset: 3D scatter plot showing all calculated properties of each spot, stoichiometry vs 2D diffusion coefficient vs polarization. Panels a bar: 1 µm

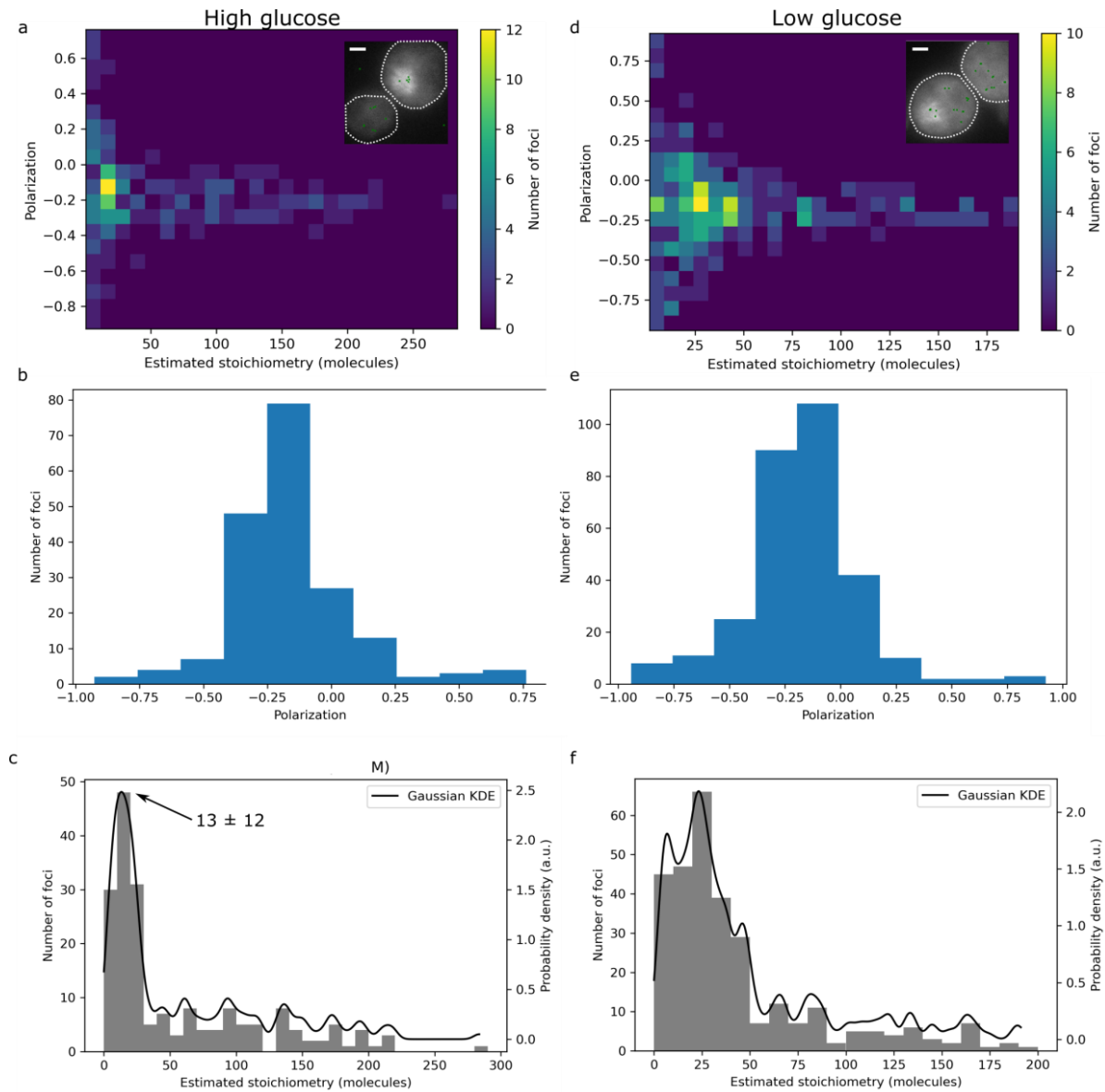

Figure 6: a) polarization vs stoichiometry 2D heatmap for yeast Mig1-GFP in 4% glucose with inset representative field of view. Cell outlines are given as dotted white lines and tracked foci as green crosses; polarization distribution for Mig1-GFP in high glucose conditions showing similar behavior to freely diffusing mGFP; c) stoichiometry distribution for Mig1-GFP aggregates in high glucose with a broad peak at 13 molecules; d) polarization vs stoichiometry heatmap for Mig1-GFP in 0.2% glucose with inset field of view as in panel a; e) polarization distribution for Mig1-GFP in low glucose; f) stoichiometry distribution for Mig1-GFP aggregates in low glucose with stoichiometry peak shifted higher, though with a higher standard deviation. N=30 cells for each dataset. Panels a and d bars: 1 μm.